\documentstyle[aps]{revtex}
\begin{document}
\draft
\title{ New exact  fronts for the nonlinear diffusion equation
with quintic nonlinearities}
\author{R.\ D.\ Benguria and M.\ C.\ Depassier}
\address{        Facultad de F\'\i sica\\
	P. Universidad Cat\'olica de Chile\\
	       Casilla 306, Santiago 22, Chile}

\date{\today}
\maketitle
\begin{abstract}
We consider travelling wave solutions of the reaction diffusion equation with
quintic nonlinearities $u_t = u_{xx} + \mu u (1 -u ) ( 1 +\alpha u + \beta u^2
+
\gamma u^3)$. If the parameters $\alpha , \beta$ and $\gamma$ obey a special
relation, then the criterion for the existence of a strong heteroclinic
connection can be expressed in terms of two of these parameters. If an
additional restriction is imposed, explicit front solutions can be obtained.
The
approach used can be extended to polynomials whose highest degree is odd.
\end{abstract}


\section{Introduction}

The nonlinear diffusion equation $u_t = u_{xx} + f(u)$ models phenomena in
 diverse
fields such as population growth, kinetics of phase transitions, chemical
reactions and many others. Of special interest is the case when the function
$f$
is such that there exist two steady states, one stable and one unstable. We
shall assume that the equation has been scaled so that
the unstable state is $u_u = 0$ and the stable state is $u_s = 1$,
and  we consider functions  $f$ which are positive  in $(0,1)$.  Then
sufficiently localized initial conditions evolve into a travelling front which
joins the two steady states\cite{AW78}. The speed at which the front
propagates, $c^*$ is
equal or greater than the linear marginal stability value $c_L = 2 \sqrt{
f'(0)}$.
 In many cases the asymptotic speed of propagation is exactly
the linear value $c_L = 2 \sqrt{f'(0)}$ obtained by the linear marginal
 stability criteria \cite{KPP37,DL83}.  There are cases however when the front
propagates at
 a speed greater than this value, case which is referred to  as that in which
a nonlinear
 speed selection mechanism \cite{BBDKL85,VS88,VS89} operates.
 Explicit expressions for this special nonlinear
 front or strong heteroclinic connection and its speed have been obtained
 for particular choices of $f$. All the known solutions
correspond to functions $f$ of the form $f(u) = \mu u + u^n -
u^{2n-1}$ which, for $\mu$ positive but smaller than a critical value $\mu_c$,
are strongly heteroclinic \cite{PT92}.
The purpose of this article is to show, using as an example a quintic
polynomial
$f$,  that the criterion for the existence
of special fronts can be formulated in many cases in a simpler way that enables
one to decide whether for a certain $f$ there is a strong heteroclinic
connection even if the exact solution for the front is not known.
 We find new exact fronts for this  quintic
polynomial for $f$ together with a criterion for strong
heteroclinicity in terms of the parameters of the polynomial
valid even when no explicit solution for the front can be obtained.
Similar results can be obtained for polynomials whose highest degree is odd.
The knowledge of exact solutions is of interest not only as a curiosity,
they  are
also needed in the framework of the recent proposal of structural
 stability\cite{PCGO94}, the knowledge of the speed for a specific form of $f$
enables the calculation of the speed for small perturbations to $f$ using
renormalization group techniques.

In section 2 we state the problem and reformulate  already
known results, and in section 3 we give the new results for the quintic
polynomial.

\section{Monotonic Fronts of the Reaction Diffusion Equation}

We consider the reaction diffusion equation
\[
u_t = u_{xx} + f(u)
\]
with $f(0) = 0$, $f(1) = 0$, $f'(0) > 0$ and $f>0$ in  $(0,1)$.
Given these conditions on $f$ then there exist fronts that connect the unstable
fixed point $u=0$ to the stable fixed point $u=1$.
  Travelling wave fronts  $u(x-ct)$
satisfy the ordinary differential equation
\begin{equation}
u_{zz} + c u_z + f(u) = 0
\qquad \lim_{z \rightarrow -\infty }u = 1, \quad
\lim_{z \rightarrow \infty }u = 0,
\end{equation}
where $z = x -c t$ and we  assume that  $c$ is positive.
 A  front joining the stable fixed
point $1$ to the unstable point $0$  is monotonic if in addition
its derivative $du/dz$ does not change sign.  If we search for monotonic fronts
 it
is convenient to consider the dependence of $z$ as a function of $u$,
or rather the dependence of
$v(u) = - (dz/du)^{-1} $
as a function of $u$. For a monotonic solution of equation (1),
$u(z)$  decreases monotonically as $z$ goes from $-\infty$ to
$\infty$, therefore the function $v(u)$ is well defined and
is positive between $1$ and
$0$ and vanishes at the fixed points.
One readily finds that the equation for $v(u)$ is
\begin{mathletters}
\label{v}
\begin{equation}
v(u) \frac{d v}{du} - c v(u) + f(u) = 0 ,
\end{equation}
with
\begin{equation}
v(0) = v(1) = 0 , \qquad {\rm with}\,\, v > 0.
\end{equation}
\end{mathletters}
 Since the endpoints are singular we must determine the
behavior near them analytically. If we consider functions $f$ analytic around
$0$ and with $f'(0) > 0$, then
near $u=0$ we find
\[
v(u) = a_1 u + a_{3/2} u^{3/2} + a_2 u^2 + a_{5/2} u^{5/2} + a_3 u^3 + \ldots
\]
where the first terms are given by
\begin{mathletters}
\begin{equation}
a_1^2 - c a_1 + f'(0) = 0 \label{eq:a1}
\end{equation}
\begin{equation}
a_{3/2} ( {5\over 2} a_1 - c) = 0 \label{eq:half1}
\end{equation}
\begin{equation}
a_2 ( 3 a_1 - c) + {1\over 2} f''(0) = 0 \label{eq:a2}
\end{equation}
\begin{equation}
a_{5/2} ( c - {7\over 2} a_1) - {7\over 2} a_{3/2} a_2 = 0 \label{eq:half2}
\end{equation}
\end{mathletters}
and so on.
That the leading term in the expansion of $v$ near zero is linear in $u$
is due  to the fact that the front in the original coordinates $u(z)$
approaches
the fixed points exponentially.
Since $v$ must be positive between 0 and 1, $a_1$ must be real and positive.
The two roots for $a_1$ are given by
 $a_{1P} = (c + \sqrt{c^2 - 4 f'(0)}\,)/2$ and
$a_{1M} = (c - \sqrt{c^2 - 4 f'(0)}\,)/2$. The minimum speed  $c$ for which
there
may be a monotonic front is the
linear marginal speed value $c_L = 2 \sqrt{f'(0)}$ value at which the roots
coincide $a_{1P} = a_{1M} \equiv a_{1L}$. For speeds greater than this value
$a_{1M} < a_{1L} < a_{1P}$. Strong heteroclinic solutions or special
nonlinear front
profiles are those associated with $a_{1P}$. From the expansion at the origin
it follows from  (\ref{eq:half1})  that
either $c = 5 a_1/2$ or $a_{3/2} = 0$. In the first case we find that
$c = 5 \sqrt{f'(0)/6} \approx 2.041 \sqrt{f'(0)}$
and $a_1 = \sqrt{2 f'(0)/3} = a_{1M}$. As it is known, these solutions
are not a preferred asymptotic state.
Strong heteroclinic connections can be achieved only if $a_{3/2} = 0$, all half
integer coefficients vanish then and $ v(u) = a_1 u + a_2 u^2 + \ldots$.

Near $u=1$, assuming $f'(1) < 0$,
\[
v(1-u) = b_1 (1-u) + b_2 (1-u)^2 + b_3 (1-u)^3 + \ldots
\]
where $b_1$ is the positive solution of
\[
b_1^2 + c b_1 + f'(1) = 0.
\]
There is only one positive solution for $b_1$, the rest of the coefficients
follow easily.

It is convenient to introduce a new parameter $\lambda$ defined by
$
c = \lambda a_1 $.
It is not difficult to realize that whenever $1 < \lambda < 2$
then the solution for $v$ is strongly heteroclinic, that is, associated with
$a_{1P}$ and when $\lambda > 2$  it becomes associated with $a_{1M}$;
 hence for $\lambda >2$ the linear
marginal speed is selected.  If $c = \lambda a_1$ then
\begin{equation}
c = \lambda \sqrt{\frac{ f'(0)}{\lambda -1}}, \qquad a_1 =
\sqrt{\frac{f'(0)}{\lambda -1}}.
\label{clue}
\end{equation}
At $\lambda = 2$, the speed attains its linear value $c_L = 2\sqrt{f'(0)}$.
The problem then is to determine the value of $\lambda$.
This transition value $\lambda=2$ is not associated with any specific
nonlinearity, it is valid for any $f$ which satisfies the conditions given
above.

 All exact front solutions given in the literature,
\cite{BBDKL85,KAL84,OPL88,VS89,PT92}
 correspond to functions $f$ for which an exact solution for  $v$ is of
the form
\[
v_n(u) = a_1 u (1-u^{n-1})
\]
which is an exact solution of equation (\ref{v})  for
\begin{equation}
f_n(u) = f'(0) \left( u + \frac{(1+n-\lambda )}{\lambda -1} u^n -
	   \frac{n}{\lambda-1} u^{2n-1} \right)
\label{super}
\end{equation}
We observe that for $\lambda = n+1$ we recover the solutions of Kaliappan
\cite{KAL84};
since $n>1$, $\lambda$ is greater than 2, so none of them are strongly
heteroclinic.
The front  corresponding to $v_n$ is given implicitly by
\begin{equation}
z = - \int \frac{du}{v_n(u)}. \label{invert}
\end{equation}
and explicitly by \cite{PT92,HMO92}
\[
u_n(z) = \frac{{\rm e}^{- z a_1}}{(1 + {\rm e}^{-(n-1) z a_1})^{1\over{n-1}} }
\]

The criterion for the existence of strongly heteroclinic fronts together with
their exact expression has been given\cite{PT92}
for functions $f$ of the form
$
f(u) = \mu u + u^n - u^{2n-1}.
$
The critical  value for $\mu$ given in \cite{PT92} for the transition from a
strong heteroclinic connection to a simple nongeneric connection (a solution
associated with $a_{1M}$) is equivalent to the value $\lambda=2$ after suitable
rescaling. It is perhaps convenient to see it in the example given by Van
Saarloos\cite{VS89}
\[
\phi_t = \phi_{xx} + \phi + d \phi^3 - \phi^5
\]
which has a strongly heteroclinic connection for $d> 2/\sqrt{3}$, of speed
\[
v^{\dag} = \frac{ - 2 + 2 \sqrt{4 + d^2}}{\sqrt{3}}.
\]

To identify the value of $\lambda$ from equation (\ref{super})
we must scale the equation for $\phi$ so that the stable
point is at 1. To do so we let $u = K \phi$,  $u$ satisfies $u_t = u_{xx} + u
+ K^2 d u^3 - K^4 u^5$ and  the stable state
is $u=1$ if
\[
K^2 = \frac{d + \sqrt{4 + d^2}}{2}
\]
where the positive sign is chosen to obtain a real $K$. Now we compare with
equation (\ref{super}) with $n=3$. We see that $f'(0) = 1$ and that
\[
 \lambda = 4 - \frac{3 d}{K^2}
\]
or, in terms of $d$,
\[
\lambda = \frac{4 \sqrt{4 + d^2} - 2 d} {d + \sqrt{4 + d^2}}.
\]
It is straightforward to see that the critical value $d = 2/\sqrt{3}$ is
exactly
$\lambda = 2$ and that the speed
\[
c = \frac{\lambda}{\sqrt{\lambda - 1}} = v^{\dag}.
\]

\section{New Solutions}

Now consider fronts for $f$ being a quintic polynomial in $u$. This problem
was considered in \cite{OPL88} but no explicit solutions were found and no
attempt to examine the conditions for the transition from the linear to the
nonlinear regime were made. Here we
show under which conditions a closed form can be obtained, together with some
examples and the condition for strong heteroclinicity $\lambda < 2$
 in terms of the
parameters of the function $f$. Evidently the value of the parameter $\lambda$
can be determined analytically only if an exact solution for $v$ is known. The
most
general form of  a quintic polynomial that vanishes at 0 and 1 is
\begin{equation}
f(x) = \mu x (1-x) (1 + \alpha x + \beta x^2 + \gamma x^3 )
\label{f5}
\end{equation}
where $\mu, \alpha , \beta $ and $\gamma$ are four arbitrary parameters whose
only
restriction is given by the requirement $f'(0)>0$ and  $f>0$ in $(0,1)$.
On the other hand, the most general closed form solution for $v$ given a
quintic
$f$ is given by
\begin{equation}
v(u) = a_1 u (1-u) (1 + b u) \label{quintic}
\end{equation}
where $b > -1$. Introducing again the parameter $\lambda$ given above, so that
$a_1$ and $c$ are given by equation (\ref{clue}), equation (\ref{quintic}) is
the exact solution of equation (\ref{v}) with
\begin{equation}
f (u) = f'(0) u (1 -u) \left( 1 + \frac{(2+\lambda b - 3 b)}{\lambda -1} u +
\frac{b(5-2b)}{\lambda-1} u^2 + \frac{3 b^2}{\lambda -1} u^4 \right).
\label{f5ex}
\end{equation}
 In the solution for $v$ we
 have three  adjustable parameters, $\lambda, b, f'(0)$
 whereas in the most general form for $f$,
four adjustable parameters exist.
Hence,  an exact
solution for $v$ can be found chosing three parameters of $f$ arbitrarily
and the  fourth one in terms
of them. Choosing $\mu, \beta$ and $\gamma$ arbitrarily, we identify
\[
f'(0) = \mu
\]
\[
\lambda = 1 + \frac{75\gamma}{(3\beta+2\gamma)^2}
\]
and
\[
b = \frac{5 \gamma}{3\beta + 2 \gamma}
\]
and the exact solution exists if
\[
\alpha = \frac{(2+\lambda b -3 b)}{\lambda -1}
\]
For any other value of $\alpha$ a closed form solution does not exist and
we cannot determine the value of $\lambda$.
 The
criterion for the solution to be strongly heteroclinic $1 <\lambda < 2$ is
expressed now in terms of the free parameters $\beta$ and $\gamma$.

 Now we show that an
explicit solution for the front in the original coordinates exists only if an
additional condition on $b$, hence a relation between the free parameters
$\beta$ and $\gamma$  is satisfied.
Proceeding as above in equation (\ref{invert}) we find that $u(z)$ is the
solution of
\begin{equation}
e^{-(b+1)\,a_1\,z} = \frac{ u^{1+b}}{(1-u) (1 + b u)^b }.
\end{equation}
Writing $b = n/p$ the equation for $u$ is
\begin{equation}
e^{-(n+p)\,a_1\,z} = \frac{ u^{n+p}}{(1-u)^p (1 + b u)^n }.
\end{equation}
This can be inverted to obtain the explicit solution for $u(z)$ if
$n+p = 2,3,4$.
The detailed inversion of all the solvable cases is not instructive, here we
give one example. Choose $n=2, p=1$, then $b=2$, and the front is a solution
of the cubic equation
\begin{equation}
u^3 ( 1 + 4 {\rm e}^{-3 a_1 z} ) - 3 u {\rm e}^{-3 a_1 z} -{\rm e}^{-3 a_1 z} =
0.
\end{equation}
This cubic has two complex roots and a single real positive root which is the
desired front, given by
\begin{equation}
u(z) =
{{{2^{{1\over 3}}}}\over
    {{\sqrt{4 + {e^{3\,a_1\,z}}}}\,
      {{\left( {e^
	     {{{3\,a_1\,z}\over 2}}} +
	   {\sqrt{4 + {e^{3\,a_1\,z}}}} \right)
	   }^{{1\over 3}}}}} +
  {{{{\left( {e^{{{3\,a_1\,z}\over 2}}} +
	  {\sqrt{4 + {e^{3\,a_1\,z}}}} \right) }
      ^{{1\over 3}}}}\over
    {{2^{{1\over 3}}}\,
      {\sqrt{4 + {e^{3\,a_1\,z}}}}}}
\end{equation}

Again this is an exact front for $f$ of the form given by equation (\ref{f5}).
It corresponds to a strongly heteroclinic connection for $\lambda < 2$.
If one chooses the case $n+p =4$ the quartic equation that arises has a pair of
complex conjugate solutions, a negative solution and a positive solution which
is the desired front.
For values of $b$ which do not allow the obtention of the explicit form of the
front $u(z)$ we still have the speed selection criteria in terms of the
free parameters of the polynomial.

Closed form solutions $v(u)$
 for polynomial $f's$ can
be obtained only if $f$ is an odd polynomial. In general, if $f$ is a
polynomial
of degree $2k +1$ that vanishes at $0$ and $1$, there are $2 k$ free parameters
(restricted only by the requirement of positivity of $f$), whereas the
corresponding closed form solution for $v$ has $k+1$ parameters, which implies
that a closed form for $v$, and an explicit expression for $\lambda$
 is possible if $k-1$ parameters of $f$ are
chosen  adequately in terms of the $k+1$ remaining free parameters.

\section{Conclusion}

We have studied the existence of exact strongly heteroclinic fronts for the
reaction diffusion with quintic nonlinearities. We find that the use of phase
space enables one to characterize the transition from strongly heteroclinic to
simple nongeneric fronts in terms of a single parameter $\lambda$ which is the
ratio between the speed and the rate of decay at infinity. The introduction of
this parameter gives a unified way in which to describe the type of
solutions which is independent of the nature of the nonlinearities. The
exact value of this parameter cannot  be determined analytically
 when the highest
nonlinearity is even, if the highest derivatives are odd it can be determined
for special choices of parameters.
In the case studied here, quintic nonlinearities, the value
of $\lambda$ can be determined exactly if a special relation betwen the
parameters of the equation is satisfied. It is not necessary to know the exact
solution $u(x-ct)$ in order to determine whether a strong heteroclinic
connection exists. If an additional restriction on the parameters is imposed,
new exact solutions can be found. We have illustrated this situation for one
particular choice, a whole family of exact solutions can be constructed.
The use of phase space is not only useful as an aid to find exact solutions, it
can be used to obtain a lower bound on the speed, valid for all $f$,
 which allows
one to determine the range of parameters for which strongly heteroclinic
connections exist \cite{BD94}.

\section{Acknowledgments}

This work has been partially supported by Fondecyt project 1930559.

\end{document}